\documentclass[12pt,preprint]{aastex}
\shorttitle{X-Ray--Emitting Ejecta Knots in Puppis~A}
\shortauthors{Katsuda et al.}

\begin{document}

\title{Discovery of Fast-Moving X-Ray--Emitting Ejecta Knots in the
  Oxygen-Rich Supernova Remnant Puppis~A}

\author{S. Katsuda\altaffilmark{1}, K. Mori\altaffilmark{2},
  H. Tsunemi\altaffilmark{1}, S. Park\altaffilmark{3},
  U. Hwang\altaffilmark{4,5}, D. N. Burrows\altaffilmark{3},
  J. P. Hughes\altaffilmark{6}, and P. O. Slane\altaffilmark{7}}

\altaffiltext{1}{Department of Earth and Space Science, Graduate School
of Science, Osaka University,\\ 1-1 Machikaneyama, Toyonaka, Osaka,
60-0043, Japan; katsuda@ess.sci.osaka-u.ac.jp}

\altaffiltext{2}{Department of Applied Physics, Faculty of Engineering,
University of Miyazaki, 1-1 Gakuen Kibana-dai Nishi, Miyazaki, 889-2192,
Japan}

\altaffiltext{3}{Department of Astronomy and Astrophysics, Pennsylvania
        State University, 525, Davey Laboratory, University Park, PA 16802
}
\altaffiltext{4}{NASA Goddard Space Flight Center, Code 662, Greenbelt
        MD 20771
}
\altaffiltext{5}{Department of Physics and Astronomy, The Johns Hopkins
        University, 3400 Charles Street, Baltimore, MD 21218
}
\altaffiltext{6}{Department of Physics and Astronomy, Rutgers
        University, 136 Frelinghurysen Road, Piscataway, NJ 08854-8019
}
\altaffiltext{7}{Harvard-Smithsonian Center for Astrophysics, 60 Garden
        Street, Cambridge, MA 02138, USA
}

\begin{abstract}

We report on the discovery of fast-moving X-ray--emitting ejecta knots
in the Galactic Oxygen-rich supernova remnant Puppis~A from {\it
XMM-Newton} observations.  We find an X-ray knotty feature positionally
coincident with an O-rich fast-moving optical filament with
blue-shifted line emission located in the northeast of Puppis~A. We
extract spectra from northern and southern regions of the 
feature.  Applying a one-component non-equilibrium ionization model
for the two spectra, we find high metal abundances relative to the
solar values in both spectra.  This fact clearly shows that the 
feature originates from metal-rich ejecta.  In addition, we find that
line emission in the two regions is blue-shifted.  The Doppler
velocities derived in the two regions are different with each other,
suggesting that the knotty feature consists of two knots that
are close to each other along the line of sight. Since fast-moving
O-rich optical knots/filaments are believed to be recoiled metal-rich
ejecta, expelled to the opposite direction against the high-velocity
central compact object, we propose that the ejecta knots disclosed
here are also part of the recoiled material.

\end{abstract}

\section{Introduction}

Puppis~A is categorized as an ``oxygen-rich'' supernova remnant (SNR)
based on optical spectroscopy (Winkler \& Kirshner 1985). The
group of O-rich remnants contains only two
other members in our Galaxy, Cassiopeia~A (Chevalier \& Kirshner 1979)
and G292.0+1.8 (Goss et al.\ 1979), and a few in the Magellanic Clouds
(Blair et al.\ 2000 and references there in).  At optical wavelengths, these 
SNRs show fast-moving metal-rich ejecta knots ($v >
1000\,\mathrm{km\,sec^{-1}}$) which are typically enriched in O and
Ne.  High-Z elements, like Ar, Ca, and Fe, are dominantly 
generated in a Type-{\scshape I}a supernova (Nomoto et al.\
1984), while low-Z elements, like O, Ne, and Mg, are found in
core-collapse SN debris (Thielemann et al.\ 1996). These facts
suggest that O-rich SNRs are core-collapse origin. The detection of
metal-rich ejecta from these SNRs thus provides us with an opportunity
to make a direct test of core-collapse SN nucleosynthesis models.

O-rich fast-moving optical knots (hereafter, OFMKs) in Puppis~A, whose
proper motions are all consistent with 
undecerelated expansion from a common center (Winkler \& Kirshner
1985; Winkler et al.\ 1988), have been found only in the northeastern
quadrant, suggesting asymmetric mass ejection during SN explosion
which produced Puppis~A.  On the other hand, proper motion of a
central compact object (CCO), RX J0822-4300, associated with the
Puppis~A SNR was recently measured to be directed toward the southwest
with a high velocity of $\sim$1600\,km\,sec$^{-1}$ (Winkler \& Petre
2007).  The CCO is considered to be kicked by asymmetric mass ejection
during the SN explosion according to momentum conservation, from which
we can study SN explosion mechanisms (e.g., Scheck et al.\ 2006). 

The so-called ``omega'' filament (Winkler \& Kirshner 1985) is one of
the best studied OFMKs in Puppis~A.
Winkler \& Kirshner (1985) observed the ``omega'' filament and found
that the spectrum was dominated by O lines, with rather weak Balmer
lines.  The mass ratio of O to H was estimated to be 30, i.e., $\sim$2000
times the solar value.  Furthermore, they found that the O lines were
blue-shifted by 1500$\pm100$\,km\,sec$^{-1}$.  These facts led
them to consider that the ``omega'' filament was nearly pure O ejecta
from the core of the SN progenitor that had remained more or less
intact.  

In X-ray wavelengths, Puppis~A generally showed sub-solar metal
abundances, suggesting that the X-ray emission from Puppis~A was
dominated by the swept-up interstellar medium (Tamura 1995; Hwang et
al.\ 2008). The complex interstellar environment such as a gradient in
the ambient density (Petre et al.\ 1982) as well as the relatively old
age of $\sim$4000 years (Blair et al.\ 2003) might have made it
difficult to detect clear signs of metal-rich ejecta in its spectra.
However, recent {\it Suzaku} observations revealed metal-rich ejecta 
features in the northeastern part of the remnant (Hwang et al.\ 2008).
Here, we also report of the discovery of the X-ray--emitting
fast-moving ejecta knots positionally coincident with the optical
``omega'' filament. 

\section{Observations and Data Reduction}

\begin{figure}
\includegraphics[angle=0,scale=0.8]{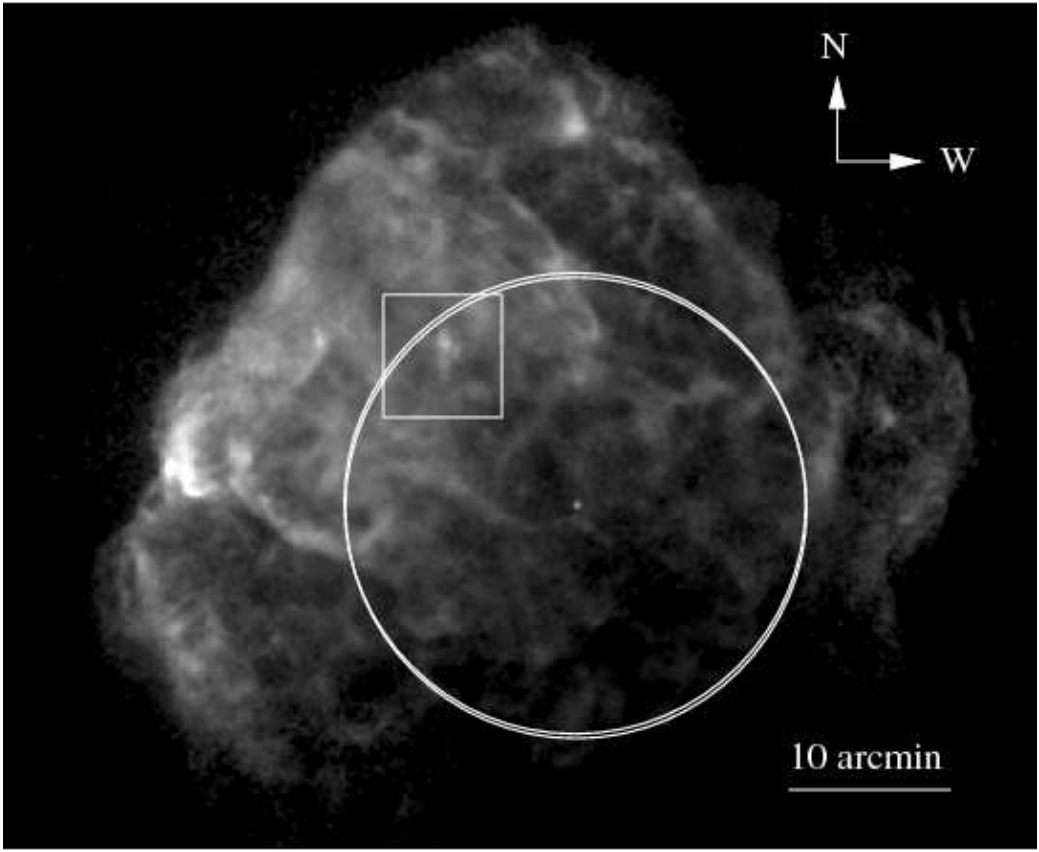}\hspace{1cm}
\caption{{\it XMM-Newton} FOV (white circles) overlaid on a {\it
  ROSAT} HRI image of the entire Puppis~A SNR.  The data have been
  smoothed by Gaussian kernel of $\sigma = 15^{\prime\prime}$.  The
  intensity scale is square root.  Optical O {\scshape III} and {\it
  XMM-Newton} three-color images in the white box region are shown in
  Fig.~\ref{fig:opt xmm}. 
} 
\label{fig:hri}
\end{figure}

Parts of Puppis~A have been observed several times with {\it XMM-Newton}.
Among them, two observations (ObsIDs.\ 0113020101 and 0113020301)
cover the ``omega'' filament (Winkler \& Kirshner 1985). The fields of
view (FOV) of the two {\it XMM-Newton} observations are shown on a
{\it ROSAT} HRI image of the entire Puppis~A in Fig.~\ref{fig:hri}.
All the raw data were processed 
with version 6.5.0 of the XMM Science Analysis Software.  We use only
MOS data, since the pn data were obtained in PrimeSmallWindow mode so
that the pn FOV covered only a small region around the CCO that is not
focused in this paper.  We select X-ray events corresponding to
patterns 0--12.  We further 
clean the data by rejecting high background (BG) intervals and
remove all the events in bad columns listed in Kirsch~(2006).  The
summed exposures from the good time intervals (GTIs) after data cleaning
are 22\,ks for ObsID.\ 0113020101 and 10.8\,ks for ObsID.\ 0113020301,
respectively.  After the filtering, the data were vignetting-corrected
using the sas task {\tt evigweight}.  

\section{Analysis and Results}

\begin{figure}
\includegraphics[angle=0,scale=0.8]{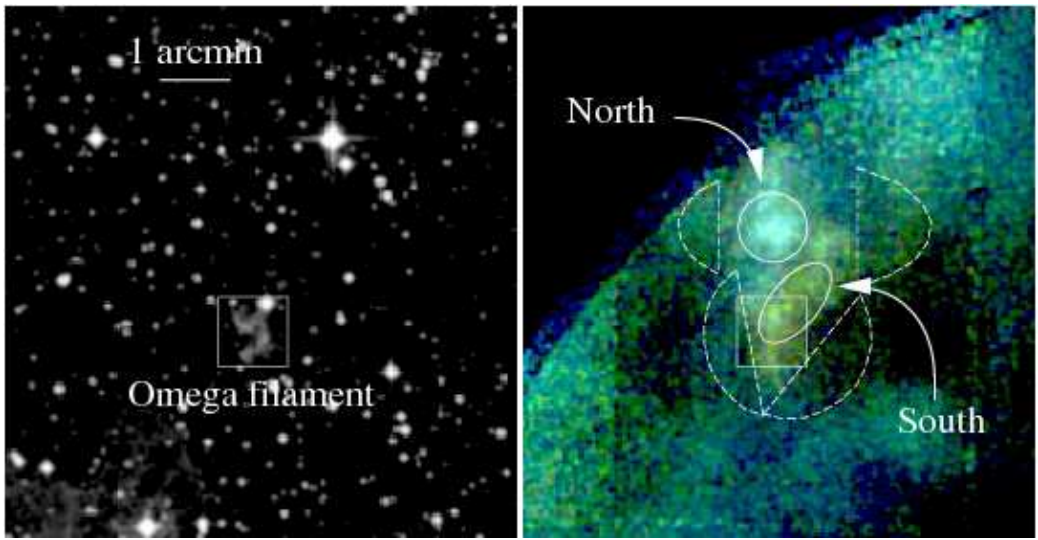}
\caption{{\it Left}: Linearly-scaled Optical O {\scshape III} image
  for the box region 
  in Fig.~\ref{fig:hri}.  The central $\Omega$-shaped filament
  enclosed in a box region is a so-called ``omega'' filament (Winkler
  \& Kirshner 1985).  {\it Right}: {\it XMM-Newton} three-color image
  in the same area.  Red, green, and blue represent 0.4--0.7,
  0.7--1.2, and 1.2-5.0\,keV, respectively.  The intensity scale is
  square root.  Spectral extraction regions are shown as solid (for 
  source spectra) and dashed (for BG spectra) white lines.  The
  location of the ``omega'' filament is indicated as a box
  region that is the same box in Fig.~\ref{fig:opt xmm} {\it Left}.}  
\label{fig:opt xmm}
\end{figure}

Fig.~\ref{fig:opt xmm} {\it Left} shows the optical O {\scshape III}
image.  We can see an $\Omega$-shaped filament, so-called ``omega''
filament in the central portion of the figure.  Fig.~\ref{fig:opt xmm}
{\it Right} shows an X-ray three-color image of the
merged MOS1/2 data which covers the same area.  Red, green, and blue
represent 0.4--0.7 (for O lines), 0.7--1.2 (for Ne lines), and
1.2--5.0 (for hard band)\,keV, respectively. We find an
X-ray--emitting knotty feature in the central 
portion of Fig.~\ref{fig:opt xmm} {\it Right}.  The southern region
of the feature is positionally coincident with the optical ``omega''
filament as indicated by box regions in Fig.~\ref{fig:opt xmm}.  

We extract spectra from the north and south of the X-ray knotty feature
because it is divided into two regions in terms of color: 
the NE region of the feature shows white color indicating a significant
mixture of spectrally hard emission, while the SW region of the feature
shows soft red emission.  The two regions are shown as a white solid
circle (radius of 30$^{\prime\prime}$) and a white solid ellipse
(angular sizes of x and y-axis of 20$^{\prime\prime}$ and
40$^{\prime\prime}$, respectively) in Fig.~\ref{fig:opt xmm} {\it
  Right}.  Note that the southern region in the feature includes the
``omega'' filament.  We simultaneously fit the MOS1 and MOS2 spectra,
allowing the normalizations between the two detectors to vary by
introducing the constant model in XSPEC (Arnaud 1996).
As BG, we subtract the X-ray emission in their surrounding regions
(i.e., areas enclosed by dashed lines around each region shown in
Fig.~\ref{fig:opt xmm} {\it Right}) after normalizing the intensities at the
ratios of source areas to  BG areas.  In this way, we attempt to
obtain the spectra of the knotty feature itself.  We note that our
method is not unprecedented; the same method was successfully
performed in analyses of ejecta knots in Cas~A SNR (Hwang \& Laming
2003; Laming \& Hwang 2003).  The count rates per area of the source
and BG regions are summarized in Table~\ref{tab:rate}.

We apply an absorbed single component non-equilibrium ionization
(NEI) model for the two spectra (the {\tt wabs} (Morrison \& McCammon
1983) and {\tt vpshock} model (NEI version 2.0) (e.g., Borkowski et
al.\ 2001) in XSPEC v\,12.3.1).  Initially, we allow the individual
element abundances to be freely fitted relative to H and find that
enhanced values are required\footnote{In this fitting, we obtain
  only lower limits of metal abundances that are typically several
  hundred times the solar values.}.  This fact
clearly shows that the knotty feature consists of metal-rich
ejecta.  The northern region is rich in O, Ne, Mg, Si, S, and Fe,
whereas the southern region is rich in O, Ne, and Mg.  We note that
Hwang et al.\ (2008) independently detect metal-abundance enhancements
positionally coincident with the X-ray knotty feature disclosed here
from their {\it Suzaku} observations, although the relatively low
spatial resolving power of {\it Suzaku} prevented them from
identifying the metal-abundance enhancements with the ``omega'' filament.
We then set O/H at the optically determined value of 2000 (Winkler \&
Kirshner 1985) in order to obtain element abundances relative to O.
Free parameters are hydrogen column density,  
$N_\mathrm{H}$; electron temperature, $kT_\mathrm{e}$; ionization
timescale, $n_{\rm e}t$; volume emission measure (VEM; VEM $=\int
n_\mathrm{e}n_\mathrm{H} dV$, where $n_\mathrm{e}$ and $n_\mathrm{H}$
are number densities of electrons and protons, respectively and $V$ is
the X-ray--emitting volume); abundances of Ne, Mg, Si, S, Fe, and
Ni.  Above, $n_{\rm e}t$ is the electron density times the elapsed
time after shock heating and the {\tt vpshock} model assumes a range of
$n_{\rm e}t$ from zero up to a fitted maximum value. We set the
abundance of Ni equal to that of Fe.  Abundances of the other elements
are fixed to the solar values \cite{Anders1989}.  

\begin{figure}
\includegraphics[angle=0,scale=0.8]{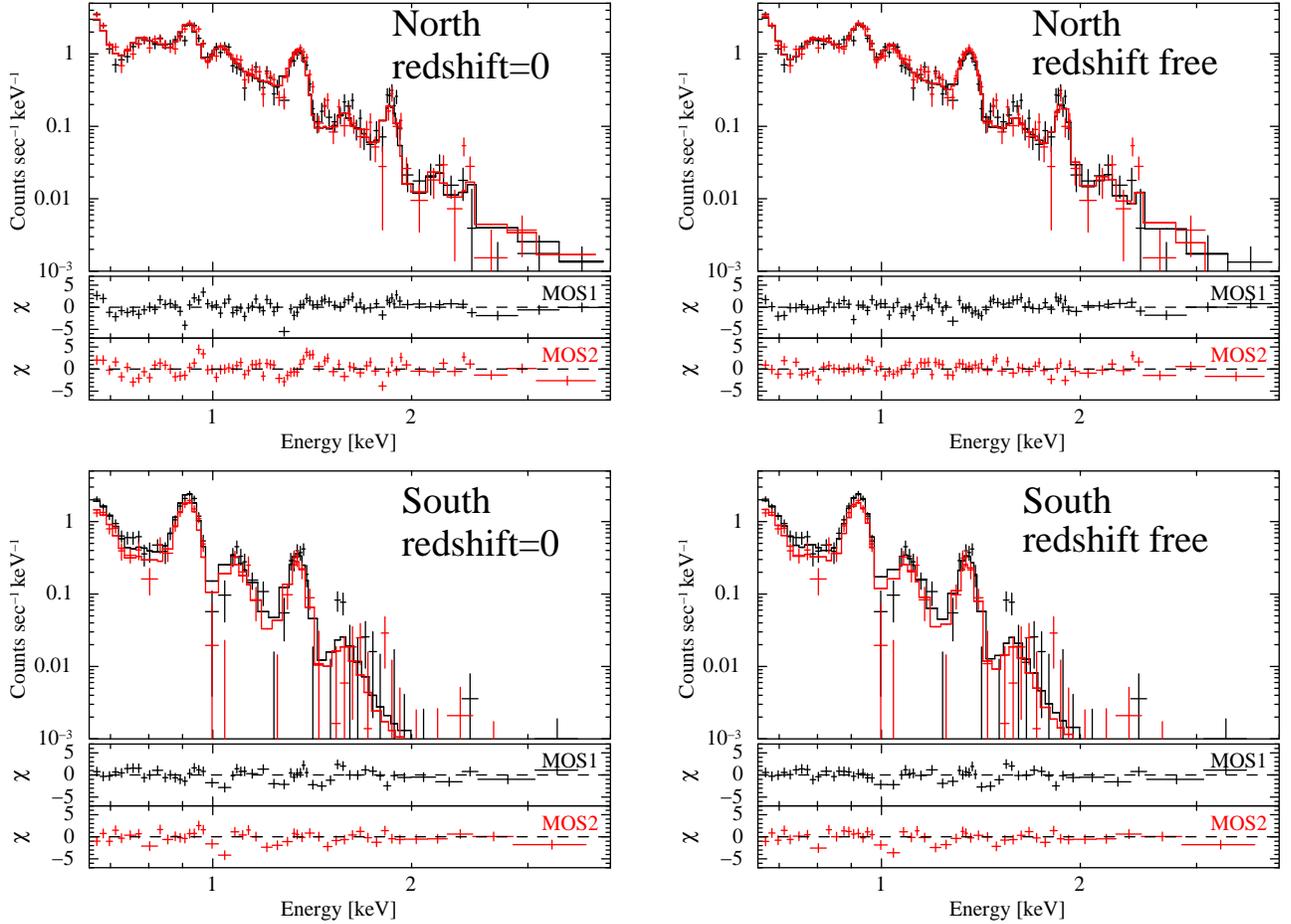}
\caption{{\it Left column}: MOS1 (black) and MOS2 (red) spectra from the two
  regions in Fig.~\ref{fig:opt xmm} with the best-fit model (redshift
  = 0).  The lower panels show the residuals.  {\it Right column}:
  Same as the left column but with the revised model (redshift is free).} 
\label{fig:ej_spec}
\end{figure}

The two spectra, along with the best-fit models, are shown in
Fig.~\ref{fig:ej_spec} (left column).  Note that we exclude data in the
energy range below 0.65\,keV where apparent differences between MOS1
and MOS2 data are seen.  The difference might be due to the
contamination on the MOS1 chip (Pradas \& Kerp 2005). The best-fit
parameters and fit statistics are summarized in
Table~\ref{tab:param}. We find that the fit level for the northern
spectrum is far from acceptable while that for the southern one is
moderate. We notice that the residuals in the northern spectrum have
wavy structures around several lines such as Ne He$\alpha$, Mg
He$\alpha$, or Si He$\alpha$. These features suggest that the line
center energies are systematically shifted from those expected by the
best-fit model.  

We thus introduce a variable redshift in the same {\tt vpshock} model.  The
two spectra with the revised best-fit model are also shown in
Fig.~\ref{fig:ej_spec} (right column).  The revised model significantly
improves the fits for both of the two spectra (with a significance
level of greater than 99.9\% based on the $F$-test).  In particular,
the wavy structures of residuals seen in the previous fit, where the
redshift is fixed to zero, for the northern spectrum are greatly reduced in
the revised fit, where the redshift is free.  Table~\ref{tab:param}
summarizes the best-fit parameters for the two spectra obtained from
the revised model fitting.  The derived redshift parameter shows
negative, i.e., blue-shifted values in the two regions.  The value of
the blueshift for the northern region of the X-ray knotty feature is higher
than that for the southern region.  We find that a relatively low
temperature and little evidence of Si and S abundances in south of the
feature, while relatively high temperatures and significant Si and S
abundances in the north of the feature. Those facts result in the
color variation seen Fig.~\ref{fig:opt xmm} {\it Right}.  Ionization
timescales for the two regions are significantly lower than that
expected for the collisional ionization equilibrium.  

Doppler velocity measurements with X-ray CCD detectors have been
successfully demonstrated by several different groups (e.g.,
Willingale et al.\ (2002) using data from {\it XMM-Newton} MOS).  However, 
small redshifts of $\lesssim 10$\,eV require careful investigation of
systematic uncertainties and/or artifacts due to the detector
calibration (the uncertainty of absolute energy scale is $\lesssim$\,5\,eV;
Kirsch 2007).  To estimate systematic errors of the derived Doppler
shifts, as well as to ensure 
that the line shifts are not due to calibration uncertainties, we
further perform spectral analysis in the following three cases: (1)
individual fit of MOS1 and MOS2 spectra, (2) individual fit of spectra
from the two observations (whose rotation angles are different from
each other), (3) inhomogeneous intensities of BG spectra, in case that the
intensities of the BG spectra in the source regions are different from
those in the BG regions currently selected.  We examine
50\% higher or lower than the original intensity since there are
inhomogeneities of X-ray intensities of $\sim$50\% around the X-ray
knotty feature.  We summarize all the derived values of the redshift in
Table~\ref{tab:uncertainty}.  We find significant blueshifts in all
the cases for the two regions. The systematic
uncertainties are greater than the statistical ones and dominate the
significance of the redshift.  The Doppler velocities, considering all
the systematic uncertainties, are
$-3400^{+1000}_{-800}$\,km\,sec$^{-1}$ for the north of the X-ray
knotty feature and $-1700^{+700}_{-800}$\,km\,sec$^{-1}$ for the south
of the feature.  The dominant uncertainty comes from the systematic
uncertainty between the two detectors and/or that introduced by
possible variations of the BG intensities.  We should note that the
values of metal abundances vary by factors of $\sim$3 in the three
cases of BG intensities examined.  This systematic uncertainty is
considered to be conservative errors of the abundances obtained.

Next, we investigate whether or not the calibration uncertainty of the
MOS energy scale is serious in our data.  We measure line center
energies in the two regions and the local BG region for each region,
applying a phenomenological model, i.e., a thermal continuum and 14
Gaussian line profiles.  The center energy, line width, normalization
of each Gaussian line, as well as the temperature and normalization of
the thermal continuum, were treated as free parameters. The derived
line center energies for Ne He$\alpha$, Mg He$\alpha$, and Si
He$\alpha$ are summarized in Table~\ref{tab:lineE}. We confirm
that the line center energies obtained in the local-BG-subtracted
spectra are indeed significantly higher than those obtained in the
surrounding BG regions.  Furthermore, the blueshifts implied by simply
comparing the source and BG line centers are quite similar to that
found in the NEI best-fit models.  These facts strongly
support the suggestion that the line shifts seen in the ejecta feature
are not due to calibration uncertainties of the MOS energy scale.   

Finally, we investigate whether the variations of line center energies
are caused by the plasma condition (i.e., ionization states, electron
temperature) rather than Doppler shifts.  According to the NEI code in
Hughes et al.\ (2003), line center energies of the K-shell complex,
including all lines from all charge states excluding Ly$\alpha$ and
all higher energy lines are calculated to be 0.872--0.918\,keV for Ne,
1.269--1.350\,keV for Mg, and 1.744--1.862\,keV for Si.  The results in
Table~\ref{tab:lineE} show that the shifts in line centroids from the
northern region cannot be related to temperature/ionization effects,
while the effects cannot be ruled out for the south region.  However,
the agreement with optical Doppler velocity makes it plausible that we
detect blueshift in the south region.

In this context, we are convinced that the line shifts observed in
the two regions are due to neither instrumental origin nor plasma
conditions (i.e., $kT_{\mathrm e}$ and $n_{\rm e}t$) but reflect the
Doppler-shifts caused by their fast motion toward us.  In conclusion,
we observe blue-shifted line emission from the north and south of the
X-ray knotty feature with Doppler velocities of
$-3400^{+1000}_{-800}$\,km\,sec$^{-1}$ and
$-1700^{+700}_{-800}$\,km\,sec$^{-1}$, respectively.  The Doppler
velocity estimated in the south of the X-ray knotty feature, where the
optical ``omega'' filament is included, is consistent with that
measured for the optical ``omega'' filament
(1500$\pm$100\,km\,sec$^{-1}$; Winkler \& Kirshner 1985).

\section{Discussion}

We find an ejecta-dominated X-ray bright knotty feature on the
position of the optically O-rich filament discovered by Winkler \&
Kirshner (1985) and Winkler et al.\ (1988).  Due to its fast motion
toward us, we detect blue-shifted lines from the feature.
The Doppler velocity in the northern region of the feature is
higher than that in the southern region.  If they are moving with
different velocities, distances from the explosion center are also
different with each other. This means that we see two different ejecta
knots that are close to each other along the line of sight.  Follow-up
work with data of better spatial resolution (i.e., {\it Chandra} data)
might reveal clear morphological separation between the north knot and
south knot. 

\subsection{Mass and Origin of the Fast-Moving Ejecta Knots}

Under the assumption of metal abundance of O/H to be 2000 times the
solar value, the electrons are dominantly supplied by O or other
metals such as Si or S.  We assume that the value of $n_\mathrm{e}$ is
7 times that of the O ion density, $n_\mathrm{O}$, because O ions mainly
exist in the He-like or H-like ionization states at the electron
temperature and the ionization timescale obtained for 
the knots (Table~\ref{tab:param}).  Assuming that the depth of the
X-ray--emitting plasma is the same as the physical scale corresponding
to the apparent angular size of the knots (1$^\prime$) at a distance of
2.2\,kpc (Reynoso et al.\ 1995), we estimate elemental densities and
masses in each region; these are summarized in Table~\ref{tab:dens_mass}.  
We obtain the total masses in the northern and southern knots to be
$\sim$0.07\,M$_\odot$, $\sim$0.08\,M$_\odot$, respectively.

The masses of individual OFMKs seen in Cas~A and Puppis~A are
estimated to be 1$\times 10^{-4}$ M$_{\odot}$ (Raymond 1984) and
1$\times 10^{-2}$ M$_{\odot}$ (Winkler et al.\ 1988), respectively.
The masses of Fe-rich knots in Cas~A and O and Ne-rich knot in G292.0+1.8
are estimated to be 4$\times10^{-5}$--1$\times10^{-3}$\,M$_{\odot}$
(Hwang \& Laming 2003) and 1$\times 10^{-3}$ M$_{\odot}$ (Park et al.\
2004), respectively.  The masses of Vela shrapnels A (Si-rich knot)
and D (O, Ne, and Mg-rich knot) are estimated to be 5$\times10^{-3}$ and
0.1\,M$_{\odot}$, respectively (Katsuda \& Tsunemi 2006; Katsuda \&
Tsunemi 2005).  Therefore, the ejecta knots disclosed here are
categorized as the most massive ejecta knots.

We investigate the origin of these ejecta material by comparing the
observed metal abundance ratios with those of theoretical models.  We
employed models for various progenitor masses 
with initial composition of solar values. We examine two sets of
models by Rauscher et al.\ (2002) and Tominaga et al.\ (2007).
For the southern knot, we compare the relative abundances
of Ne and Mg to O with those expected in the models, and find that
some models by both of the two groups can reproduce the derived
relative abundances in the explosive Ne/C-burning cores.  On the other
hand, the composition of the northern knot turns out not to be
reproduced in any models at any mass radius.  If we compare the mass
ratios of S and Fe to Si to those of the models, we find that all the
models examined can reproduce the mass ratios in incomplete explosive
Si-burning layers.  The progenitor mass and specific burning process
are not well constrained by the current data analysis.  This is
partly because the local BG subtraction affects the abundance
determination by a factor of $\sim$3, and partly because emission 
from north and south knots are contaminating with each other in our
current data.  Follow-up observations with better spatial resolution
are needed to reveal the origins of the two knots as well as the
progenitor mass.

\subsection{Recoil to the High-Velocity CCO}

Considering that OFMKs in Puppis~A are believed to be recoiled
materials to the high-velocity CCO (Hui \& Becker 2006; Winkler \&
Petre 2007), we suggest that the ejecta knots which we find
near the OFMKs are also part of the recoiled materials. Winkler \&
Petre (2007) measured the proper motion of the stellar remnant and
estimated the momentum to be
$\sim4\times10^{41}\mathrm{g\,cm\,s^{-1}}$ in the plane of the sky.  
They also estimated approximate total
  momentum for the 11 measured OFMKs toward
the opposite direction to the traveling direction of the CCO to be
$\sim1.3\times10^{41}\mathrm{g\,cm\,s^{-1}}$, comprising about
$\sim$30\% of the required momentum to balance that of the CCO.  
We estimate that for the X-ray--emitting ejecta knots to be
$\sim0.5\times10^{41}\mathrm{g\,cm\,s^{-1}}$, comprising about 10\% of
the required momentum.  The rest of the required momentum might be
explained by unidentified $\sim$20 OFMKs whose existence was
suggested in O {\scshape III} image of this remnant (Winkler \&
Petre 2007). 

There are mainly three mechanisms proposed to explain the origin
of high-velocity neutron stars such as the CCO in Puppis~A: binary
disruptions, natal or postnatal kicks from the SN explosions (see, Lai
et al.\ 2001 for a review).  Recent theoretical studies have focused
on natal kicks imparted to neutron stars at birth rather than the
other two mechanisms.  A natal kick could
be due to global hydrodynamical perturbations in the SN core (e.g.,
Goldreich et al.\ 1996), or it could be a result from asymmetric
neutrino emission in the presence of super strong magnetic fields 
($B \gtrapprox 10^{15}$ G) in the proto--neutron star (e.g., Blandford et
al.\ 1983). The two models lead to distinct predictions: the measured
neutron star velocity should be directed opposite to the momentum of
the gaseous SN ejecta caused by linear momentum conservation in the
hydrodynamically-driven mechanisms (e.g., Scheck et al.\ 2006) while
neutrino-driven mechanisms predict the motion of the ejecta
roughly in the same direction of the neutron star kick (Fryer \&
Kusenko 2006).  Therefore, our data as well as the distributions of
OFMKs suggest that the hydrodynamically- (or ejecta-)driven mechanisms
are at work to produce the high-velocity CCO in Puppis~A.
However, we should care about the possibility that the lack of
  OFMKs in the southwestern portion of the remnant might otherwise be
  explained by complicated structures of the ambient medium (e.g.,
  Petre et al.\ 1982).  Also, the existence of ejecta knots disclosed
  here is not inherently convincing evidence for a lack of undetected
  ejecta in other directions, although diffuse X-ray emission is
  generally reported to have low metal-abundances (Tamura 1995; Hwang
  et al.\ 2008).  To obtain further conclusive evidence that the
  ejecta-driven mechanism is at work, we need detailed spatially
  resolved spectral analysis for the entire remnant.

\section{Summary}

We have analyzed archival {\it XMM-Newton} data of the Puppis~A SNR.  
We disclose an X-ray knotty feature on the position of one of OFMKs
discovered by Winkler \& Kirshner (1985) and Winkler et al.\ (1988).
We find that the X-ray knotty feature is metal-rich ejecta with
blue-shifted emission lines.  The composition in the northern region
of the feature is different from that in the southern region: O, Ne,
Mg, Si, S, Fe-rich in the northern region, whereas O, Ne, and Mg-rich
in the southern region. Also, the Doppler velocity in the northern region,
$-3400^{+1000}_{-800}$\,km\,sec$^{-1}$, is different from that in the
southern region, $-1700^{+700}_{-800}$\,km\,sec$^{-1}$.  These facts
lead us to consider that the feature consists of two different knots
that are close to each other along the line of sight.  Current data
are not sufficient to constrain the origins of the two knots as well
as to precisely determine the Doppler velocities in the two knots.
Further observations with better spatial resolution and better
observational condition, especially considering that the knots are
located at the edge of FOV in the current data, are strongly required
to reveal possible morphological separation between the north knot and
south knot, and to accurately determine the compositions as well as
Doppler velocities in the two knots.

\acknowledgments

This work is partly supported by a Grant-in-Aid for Scientific Research
by the Ministry of Education, Culture, Sports, Science and Technology
(16002004). This study is also carried out as part of the 21st Century
COE Program, \lq{\it Towards a new basic science: depth and
synthesis}\rq. The work of K.M.\ is partially supported by the
Grant-in-Aid for Young Scientists (B) of the MEXT (No.\ 18740108).
S.P.\ was supported in part by the NASA grant under the contract
NNG06GB86G. P.O.S.\ acknowledges support from NASA Contract
NAS8-39073. S.K.\ is supported by JSPS Research Fellowship for Young
Scientists.

\clearpage

\begin{deluxetable}{lcccccc}
\tabletypesize{\scriptsize}

\tablecaption{Count rates per area (counts\,sec$^{-1}$\,arcmin$^{-2}$)
  in the source and BG regions.}  
\tablewidth{0pt}
\tablehead{
\colhead{} &\colhead{North} &\colhead{North (BG)}&\colhead{South}
  &\colhead{South (BG)}
}
\startdata
MOS1 & 
2.53$\pm$0.01  & 1.40$\pm$0.01 & 1.51$\pm$0.01& 0.99$\pm$0.01 \\ 
MOS2 & 
2.42$\pm$0.01 & 1.26$\pm$0.01& 1.37$\pm$0.01& 0.99$\pm$0.01\\
\enddata

\tablecomments{Count rates are estimated in the energy range of
  0.65--4.0\,keV.} 
\label{tab:rate}
\end{deluxetable}

%\begin{flushleft}
%\oddsidemargin -0.5cm
\begin{deluxetable}{lcccccccccccc}
\tabletypesize{\tiny}
\tablecaption{Spectral-fit parameters in the two regions shown in
  Fig.~\ref{fig:opt xmm} {\it Right}}
\tablewidth{0pt}
\tablehead{
\colhead{Region}&
\colhead{$N_\mathrm{H}$}&
\colhead{$kT_\mathrm{e}$ (keV)}& \colhead{log($n_{\rm e}t$)}&
\colhead{Ne/O}&\colhead{Mg/O}& \colhead{Si/O}& \colhead{S/O} &
\colhead{Fe/O}& \colhead{VEM$^a$} &
\colhead{redshift ($\times10^{-3}$)}&\colhead{$\chi^2/d.o.f.$}
}
\startdata

North&0.37$^{+0.03}_{-0.02}$& 1.36$^{+0.11}_{-0.08}$& 10.65$^{+0.05}_{-0.05}$
& 0.8$\pm0.1$ & 0.8$\pm0.1$ & 0.5$\pm0.1$
& 0.4$\pm0.3$ & 0.22$^{+0.08}_{-0.06}$ & $1.2\pm0.2$ & --- &  462/216 \\

South&0.43$\pm0.02$& 0.36$\pm0.02$& 10.59$^{+0.06}_{-0.08}$
& 1.4$^{+0.3}_{-0.1}$ & 1.7$^{+0.5}_{-0.3}$ & $<$1.1
& $<$0.7 & $<$0.03 & 2.7$\pm0.3$ & --- &  290/191 \\

\\

North&0.40$^{+0.03}_{-0.06}$& 0.60$^{+0.08}_{-0.05}$& 11.20$^{+0.07}_{-0.06}$
& 1.04$^{+0.05}_{-0.06}$ & 0.95$^{+0.19}_{-0.05}$ & 0.7$\pm0.1$
& $<$0.8 & 0.16$\pm0.03$ & $2.2\pm0.1$ & $-11.4^{+0.4}_{-0.3}$
&  288/215 \\ 

South&0.43$^{+0.01}_{-0.02}$& 0.35$\pm0.01$& 10.71$^{+0.02}_{-0.14}$
& 1.5$\pm0.1$ & 1.7$^{+0.2}_{-0.3}$ & $<$0.5
& $<$0.7 & $<$0.03 & 2.5$\pm0.3$ & $-5.5^{+0.1}_{-1.0}$ &  266/190 \\

\enddata

\tablecomments{The best-fit parameters for the two spectra in
 Fig.~\ref{fig:opt xmm} {\it Right}.  Results are from the {\tt vpshock} model in
 which the redshift is fixed to zero (upper two rows) and allowed to vary
 (lower two rows). The units of $N_{\rm H}$ and VEM are
 $\times10^{22}$\,cm$^{-2}$ and 
 $\times10^{53}$\,cm$^{-3}$, respectively.  VEM is calculated at a
 distance of 2.2\,kpc (Reynoso et al.\ 1995).  Quoted errors are at 90\%
 confidence level. 
 The values of the abundance ratios are relative to those of the
 solar ratios.  Other elements are fixed to those of the solar values
 \cite{Anders1989}. The errors are calculated after fixing the
 $kT_\mathrm{e}$ to the best-fit value.  Errors for VEM are
 calculated after fixing the $kT_\mathrm{e}$ and Ne abundance to the
 best-fit values. 
}
\label{tab:param}
\end{deluxetable}
%\end{flushleft}

\begin{table}
\begin{center}
%\tabletypesize{\scriptsize}
\caption{Redshift values ($\times10^{-3}$)}
\begin{tabular}{lcccccccc}
\tableline\tableline
Region &\multicolumn{2}{c}{Case-1$^a$}
&\multicolumn{2}{c}{Case-2$^b$}  &
\multicolumn{3}{c}{Case-3$^c$}\\ 

& MOS1&MOS2&Obs1&Obs2&0.5&1$^d$&1.5\\
\tableline
North & $-8.6^{+0.2}_{-0.7}$ & $-12.4^{+0.6}_{-0.4}$
& $-8.6\pm0.4$ & $-12.6^{+0.7}_{-0.5}$
& $-9.0^{+0.2}_{-0.1}$ & $-11.4^{+0.4}_{-0.3}$& $-13.9^{+0.1}_{-0.2}$\\

South & $-5.6^{+1.6}_{-0.9}$ & $-5.6^{+2.2}_{-2.8}$
& $-5.1\pm0.2$ & $-6.7^{+3.5}_{-1.7}$
&$-5.7^{+0.1}_{-0.9}$ &$-5.5^{+0.1}_{-1.0}$&  $-5.5^{+2.3}_{-1.2}$ \\

\tableline
\end{tabular}
\tablecomments{Quoted errors are at 90\% confidence level. $^a$Case-1;
  we individually analyze MOS1 and MOS2 data.  In this case, we include
  both of the two observations and subtract the local BG with original
  intensity. $^b$Case-2; we separately analyze the data from the two
  observations.  Obs1 and Obs2 are ObsID.\ 0113020101 and ObsID.\ 
  0113020301, respectively. In this case, we use data from both MOS1 and MOS2
  detectors and subtract the local BGs with original intensities.
  $^c$Case-3; we artificially vary the intensities of the local BGs (0.5 and
  1.5 times the original intensity).  In this case, we use all the
  data sets (MOS1+MOS2+Obs1+Obs2).  $^d$Same as listed in
  Table~\ref{tab:param}. }  

\label{tab:uncertainty}
\end{center}
\end{table}

\begin{deluxetable}{lcccccc}
\tabletypesize{\scriptsize}

\tablecaption{Line center energies of spectra from regions shown in
  Fig.~\ref{fig:opt xmm} {\it Right}.} 
\tablewidth{0pt}
\tablehead{
\colhead{line} &\colhead{North} &\colhead{North (BG)}&\colhead{South}
  &\colhead{South (BG)}
}
\startdata
Ne He$\alpha$ (eV) & 
930$^{+6}_{-5}$  & 917$\pm$2 &922$^{+5}_{-3}$ & 917$\pm$1 \\ 
Mg He$\alpha$ (eV) & 
1357$\pm$2 &1344$^{+1}_{-2}$ &1351$^{+3}_{-5}$ & 1346$^{+1}_{-3}$\\
Si He$\alpha$ (eV) & 
1878$^{+5}_{-8}$ & 1856$^{+4}_{-3}$ &ND$^a$ & 1857$^{+3}_{-2}$\\ 
\enddata

\tablecomments{Quoted errors are at 90\% confidence level.  $^a$We
  could not determined the value due to the poor statistics.}
\label{tab:lineE}
\end{deluxetable}

\begin{deluxetable}{lccccccc|cccccc}
\tabletypesize{\scriptsize}

\tablecaption{Densities (cm$^{-3}$) and masses (M$_\odot$) in the two
  regions in Fig.~\ref{fig:opt xmm} {\it Right}}
\tablewidth{0pt}
\tablehead{
\colhead{} &\colhead{$n_\mathrm{e}$}&\colhead{$n_\mathrm{O}$} &\colhead{$n_\mathrm{Ne}$}&\colhead{$n_\mathrm{Mg}$} &\colhead{$n_\mathrm{Si}$}&\colhead{$n_\mathrm{S}$} &\colhead{$n_\mathrm{Fe}$}&\colhead{M$_\mathrm{O}$} &\colhead{M$_\mathrm{Ne}$} &\colhead{M$_\mathrm{Mg}$} &\colhead{M$_\mathrm{Si}$} &\colhead{M$_\mathrm{S}$} &\colhead{M$_\mathrm{Fe}$}
}
\startdata
North&5&0.7&0.1&0.03&0.02&0.005&0.002&0.05&0.01&0.003&0.003&0.001&0.0015\\
South&5&0.7&0.15&0.05&0&0&0&0.06&0.015&0.007&0&0&0\\
\enddata
\tablecomments{Typical errors (which mainly come from the assumptions
  of the plasma depth and the filling factor ) are about a factor of
  2.} 
\label{tab:dens_mass}
\end{deluxetable}

\end{document}